\documentclass[final,1p,times]{elsarticle}

\usepackage{graphicx}
\usepackage{amssymb}
\usepackage{amsmath}
\usepackage{siunitx}

\biboptions{sort&compress}

\journal{Physica E}

\newcommand{\vb}{\boldsymbol}
\DeclareMathOperator{\dif}{d}

\DeclareMathOperator{\cov}{cov}
\DeclareMathOperator{\PDF}{PDF}

\begin{document}

\begin{frontmatter}


\title{Multifractality of \textit{ab initio} wave functions in doped semiconductors}



\author[UF,UW]{Edoardo G.\ Carnio}
\address[UF]{Physikalisches Institut, Albert-Ludwigs-Universit\"{a}t Freiburg, Hermann-Herder-Stra{\ss}e 3, 79104 Freiburg, Germany}

\author[UW]{Nicholas D.M.\ Hine}
\address[UW]{Department of Physics, University of Warwick, Coventry CV4 7AL, United Kingdom}

\author[UW]{Rudolf A.\ R\"{o}mer}

\begin{abstract}
In Refs.\ \cite{Carnio2019,Carnio2018} we have shown how a combination of modern linear-scaling DFT, together with a subsequent use of large, effective tight-binding Hamiltonians, allows to compute multifractal wave functions yielding the critical properties of the Anderson metal-insulator transition (MIT) in doped semiconductors.  
This combination allowed us to construct large and atomistically realistic samples of sulfur-doped silicon (Si:S). The critical properties of such systems and the existence of the MIT are well known, but experimentally determined values of the critical exponent $\nu$ close to the transition have remained different from those obtained by the standard tight-binding Anderson model. In Ref.\ \cite{Carnio2019}, we found that this ``exponent puzzle'' can be resolved when using our novel \emph{ab initio} approach based on scaling of multifractal exponents in the realistic impurity band for Si:S. 
Here, after a short review of multifractality, we give details of the multifractal analysis as used in \cite{Carnio2019} and show the obtained \emph{critical} multifractal spectrum at the MIT for Si:S. 
\end{abstract}


\end{frontmatter}



\section{Introduction}\label{ch:4a}

The Anderson metal-insulator transition (MIT) \cite{Anderson1958a} is one of the fundamental manifestations 
of the quantum mechanical nature of disordered materials \cite{Kramer1993,Belitz1994,Evers2008}. In his 1958 publication \cite{Anderson1958a}, Anderson studies the localization of electrons in doped semiconductors.
The existence of the MIT in these materials was later confirmed by measuring the scaling of the conductance when increasing the dopant concentration beyond a critical value \cite{Thomas1983,Thomas1985,Stupp1993,Waffenschmidt1999}. The critical properties of the MIT, such as the exponent $\nu$ of the conductivity, should be universal quantities \cite{Cardy1996,Lopez2012}. 
For classical waves \cite{Wiersma1997,Scheffold1999,Storzer2006,Sperling2016,Skipetrov2016,Schwartz2007,Hu2008,Faez2009,Aubry2014,Hildebrand2014,Cobus2016} and cold atom systems \cite{Billy2008,Roati2008,Clement2006,Kondov2011,Jendrzejewski2012,Semeghini2015,Lopez2012} results agree well with many of the non-interacting Anderson model estimates \cite{Slevin1999,Rodriguez2010}.
However, in  experiments with semiconductors, 
$\nu$ is found to vary with sample-specific properties, namely the dopant concentration around the transition point, the homogeneity of the doping, and the purity of the sample itself \cite{ItoWOH04}. The term ``exponent puzzle'' \cite{Thomas1985,Stupp1993} was hence coined to describe this inability to characterize the Anderson transition in terms of a single, universal value for $\nu$.

Recently, we presented a study \cite{Carnio2019} that moves beyond the paradigmatic, highly-simplified, tight-binding Anderson model, and employs atomistically correct \textit{ab initio} simulations \cite{Neugebauer2013,Jain2016} of the doped semiconductor Si:S \cite{Winkler2011}.
With this approach we observe how the impurity band (IB) forms and eventually merges with the conduction band upon increasing the dopant concentration. We then exploit the multifractal nature of the (near-)critical electronic wave functions as a basis for a finite-size scaling analysis that aims to retrieve the critical properties of the transition in the thermodynamic limit.
In order to reach sufficiently large system sizes, we devised a hybrid approach: linear-scaling DFT calculations using the ONETEP code \cite{Skylaris2005}, on prototype systems of $8 \times 8 \times 8$ diamond-cubic unit cells ($4096$ atoms) to construct catalogs of local Hamiltonian blocks to describe the no/single/double dopant situation. For each concentration of impurities and disorder realization, we then built much larger, effective tight-binding Hamiltonians $H$ and overlap matrices $O$ from these catalogs for system size $L$, and solved the large generalized eigenvalue problem $H \psi_j = \epsilon_j O \psi_j$ for eigenenergies $\epsilon_j$ and eigenvectors $\psi_j(\mathbf{r})$, with $\mathbf{r}=(x,y,z)$ for coordinates $x,y,z= 1, \ldots, L$. 

In this paper, we show in detail how to perform the multifractal analysis of the $\psi_j$ and present the resulting \emph{critical} multifractal spectrum. Our results suggest that it is different from the spectrum in the Anderson model \cite{Carnio2018}. In addition, we give further details in the finite-size scaling analysis needed to ascertain the existence and the properties of the MIT when $L\rightarrow\infty$.

\section{Some of the basics of multifractals}\label{sec:mfa}

For the Anderson transition from a metal to an insulator upon increasing the disorder \cite{Anderson1958a}, the absence of length scales at criticality means that the wave function intensity $|\psi(\vb r)|^2$ at the critical disorder is \emph{self-similar} \citep{Aoki1982,Kramer1993,Evers2008}.
%
It needs to have a ``filamentary'' structure \citep{Aoki1983}, i.e.\ it needs to be extended throughout the volume, a property of the metal phase, but also to occupy only an infinitesimal fraction of it, a property of the localised phase. This structure allows the critical phase to be continuously connected to both the extended and localised phases.
In conjunction with its self-similar property, the critical wave function qualifies as a \emph{fractal} \cite{Mandelbrot1984}, at least as long as we can disregard the lower limit imposed by the lattice spacing $ a $.
Castellani \emph{et al.} \cite{Castellani1986} realised, based on the earlier work of Wegner \cite{Wegner1980}, that the critical wave function is not a simple fractal, but rather an ``interwoven family'' of fractals, each with its own dimension and distribution. Such an object is a \emph{multifractal} \citep{Mandelbrot1972,Mandelbrot1974}.

\subsection{Self-similarity}\label{sec:mfa-link}

Let us consider a system occupying a finite region of space $ \Sigma \subset \mathbb{R}^D$ with a local density $ \rho(\vb r) $. Following Ref.\ \cite{Pietronero1990}, we define the pair correlation function,
$
g(\vb r) = \langle \rho(\vb r + \vb r') \rho(\vb r') \rangle_{\vb r'}
$, 
which gives the probability that two points separated by $ \vb r $ both belong to the region $ \Sigma $. For simplicity, we now assume that the correlation function is isotropic, $ g(\vb r) = g(r) $.
In the absence of length scales, $ g $ obeys \emph{homogeneity laws} (or \emph{scale-invariance}) with respect to a \emph{resolution} or \emph{coarse-graining} $ \lambda $. More specifically, if we rescale lengths as $ r \rightarrow r' = \lambda r $ we have that
$
g(r') = \lambda^\kappa g(r) 
$, 
where $ \kappa $ is a \emph{homogeneity exponent}. The solution to this equation is given by a \emph{power-law behaviour}, $ g(r) \propto r^\kappa $. If we then fix $ r $ as the reference length scale, and $ g(r) = 1 $, we see that 
\begin{equation}\label{eq:self-similarity}
g(\lambda) = \lambda^\kappa
\end{equation}
translates the concept of self-similarity into a mathematical relation.
We can hence describe fractal objects as self-similar structures whose observed spatial extent (e.g.\ volume) depends, with a power-law behaviour, on the resolution at which we look at it. For fractals originating from a mathematical relation, the dependence on the resolution can extend over an infinite range. For fractals appearing in physical systems, instead, the range of $ \lambda $ is usually limited by macro- and/or microscopic scales. A very comprehensive list of examples can be found, e.g., in \cite{Malcai1997a}.

\subsection{Measures, fractals and multifractals}\label{sec:mfa-measures}
Let $ \psi(\vb r) $ be the wave function of an electron in a $ L \times L \times L $ volume. The modulus square $ |\psi(\vb r)|^2 $ defines a normalised measure on this volume,
which we partition in boxes of linear size $ l = \lambda L $. The number of boxes will then be $ \lambda^{-d} $, where $ d = 3 $ is the Euclidean dimension of the support of the system. The probability of finding the electron in box $ \mathcal{B}_i $ is the \emph{box-probability}
$
\mu_i = \int_{\mathcal{B}_i} |\psi(\vb r)|^2 \dif \vb r 
$. 
We can then compute the \emph{fractal dimension} $ D $ of the system by counting the number of boxes where the box-probability does not vanish:\footnote{It is customary to use $ \sim $ to indicate that the proportionality constant is independent of the resolution and can thus be ignored. This constant might appear, for instance, when the boxes, whichever their shape, do not perfectly cover the system. Since in our analysis we are covering boxes with boxes, most of the relations in this section are actually equalities.} $ N(\lambda) \sim \lambda^{-D}$. Because the electron can access any portion of the volume, i.e. there is no region of space with vanishing probability, we trivially conclude that the fractal dimension is $ D = d = 3 $.

Compared to the fractal dimension, more insightful is actually the study of the powers of the box-probability $ \mu^q_i $, which is the idea behind \emph{multifractal} analysis.
%
\emph{If} the wave function is a multifractal, we expect to see the power-law behaviour of \eqref{eq:self-similarity}:
\begin{equation}\label{eq:def-tau}
\langle \mu^q  \rangle_L \sim \lambda^{D+\tau_q} \, ,
\end{equation}
where $ \langle \ldots \rangle_L $ denotes the average over all boxes in the volume. Equivalently, we can introduce the \emph{partition sum} $ R_q(\lambda) $ (also the \emph{generalised inverse participation ratio}) as $ \langle \mu^q  \rangle_L = \lambda^D R_q(\lambda)$ and write
\begin{equation}\label{eq:Rq}
R_q(\lambda) = \sum_i \mu_i^q \sim \lambda^{\tau_q} \, .
\end{equation}
The \emph{mass exponents} $ \tau_q $ describe the scaling behaviour of the moments and do not depend on $ \lambda $.

Let us stress again that multifractality holds if, in the power-law relation of Eq.\ \eqref{eq:def-tau}, we have $\tau_q \neq 0$ for a finite range of $ \lambda $: the box size $ l $ should be smaller than the system size, but also larger than the microscopic scale $ a $, usually equal to the lattice spacing in the Anderson case. At the same time, for critical states at the Anderson transition, the system size is much smaller than the correlation length $ \xi $, such that
\begin{equation}\label{eq:length-scales}
a \ll l < L \ll \xi
\end{equation}
Additionally, the wave function is truly critical (and hence multifractal) only in the thermodynamic limit, where both $ L $ and $ \xi $ diverge. Thus $ \tau_q $ is defined in the limit $ \lambda \rightarrow 0 $. For finite systems, instead, we choose states and coarse-grainings that satisfy \eqref{eq:length-scales}. In this case, we can \emph{estimate} $ \tau_q $ by fitting the slope of $ \log R_q(\lambda) $ versus $ \log \lambda $. We are assuming here that multifractality survives in finite systems \citep{Cuevas2007}, and postpone the discussion of this non-trivial assumption to section \ref{sec-scaling}.

From \eqref{eq:def-tau} and the normalisation of the wave function, it is possible to show that $ \tau_0 = -D$ and $ \tau_1 = 0 $. This implies that we can generalise the definition of the fractal dimension to the \emph{anomalous dimensions} $ D_q $ such that $ D_0 = D $ and \emph{mass strengths} $ \tau_q = D_q \, (q-1)$. 
In the case of a simple fractal $ D_q \equiv D $, while for a multifractal $ D_q $ has a non-trivial dependence on $ q $.
The deviation from the simple-fractal case is captured by the \emph{anomalous scaling exponent} $ \Delta_q = (D_q-d)(q-1) = \tau_q - d(q-1)$.
\begin{figure*}
\includegraphics[width=\textwidth]{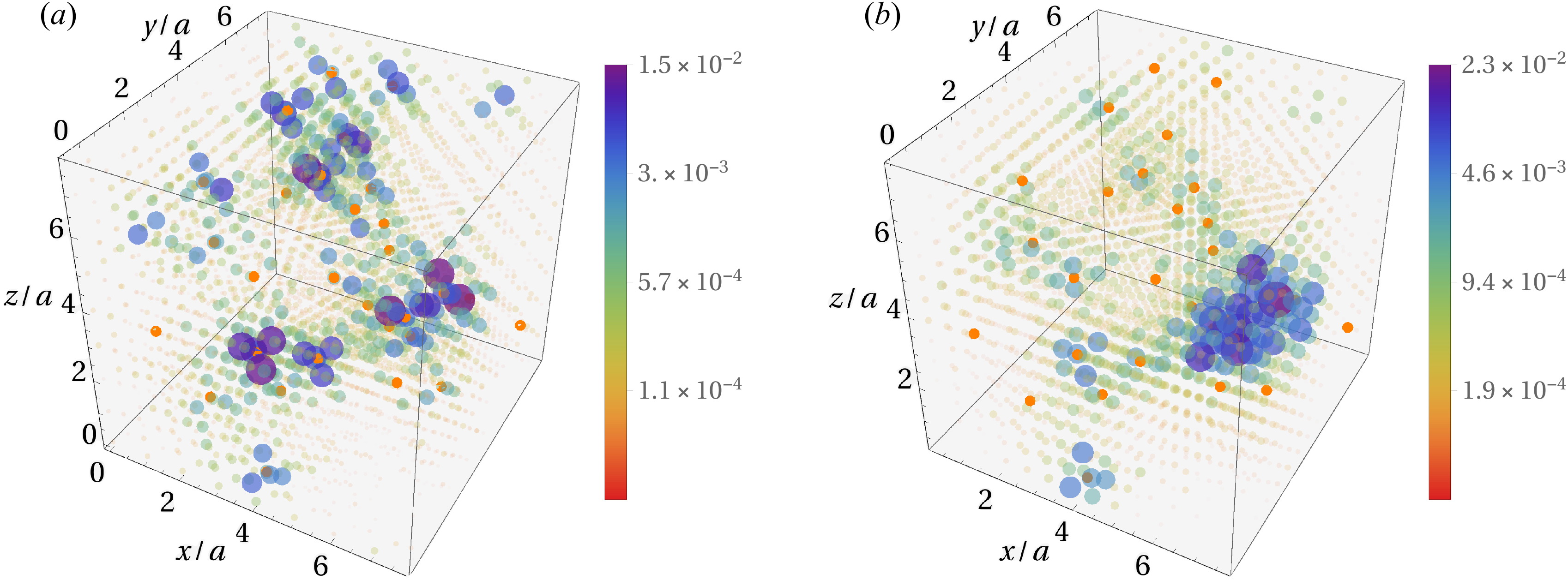}

\caption{
		Wave function of the impurity-band state closest to $\varepsilon_\text{F}$ for an exemplary systems of $ \num{4067} $ Si atoms and $ 29 $ S impurities. (\textit{a}) Shows the state calculated with ONETEP, while in (\textit{b}) calculated in the effective tight-binding model.
		%
		We have represented the top 90\% wave function values $ \psi_i $ with spheres of volume proportional to $ |\psi_i|^2 $. Opacity and colour are proportional to $ - \log_L |\psi_i|^2$, with $ L=16 $ here, so that lower (higher) values are in red transparent (violet solid). Orange dots indicate the position of the impurities. We use $ a $ to denote the Si lattice parameter. 
		\label{fig:wave-functions}}
\end{figure*}

\subsection{The multifractal spectrum}

The scaling of the moments $ R_q $, yielding $ \tau_q $, is enough to fully characterise the multifractal nature of the wave function. Now we present an equivalent description of the multifractal that will be useful, in the sections that follow, to validate our results and compare them to the 3D Anderson model.
%
This description is founded on a \emph{multifractal measure} \citep{Frisch}, a distribution such that, around each box $\mathcal{B}_i$, $ \mu_i = \lambda^{\alpha_i} $. The set of boxes with $ \alpha_i \in [\alpha,\alpha + \dif \alpha] $, then, constitutes a simple fractal with dimension $ f(\alpha) $, such that the number of said boxes is
\begin{equation}\label{eq:alpha-histogram}
N_\lambda(\alpha) \sim \lambda^{-f(\alpha)} \qquad \text{and} \qquad \alpha_i = \frac{\log \mu_i}{\log \lambda}\, .
\end{equation}
This is the formalisation of the idea of Castellani \cite{Castellani1986} that the multifractal is composed of different simple fractals.
We re-express the partition sum of Eq.\ \eqref{eq:Rq} as
\begin{equation}\label{eq:saddle}
R_q(\lambda) = \sum_i \mu^q_i = \sum_i \lambda^{q \alpha_i} = \int N(\alpha) \lambda^{q \alpha} \dif \alpha \sim \int \lambda^{q \alpha - f(\alpha)} \dif \alpha \, .
\end{equation}
For small $ \lambda $, we can use the saddle point approximation and find that the biggest contribution in the integral \eqref{eq:saddle} comes from the value of $ \alpha $ that maximises (since $ \lambda < 1 $) the argument of the exponential, i.e.\ the $ \alpha_q $ such that $ f'(\alpha_q) = q $. We can then write, from \eqref{eq:Rq}, $ \tau_q = q \alpha_q - f(\alpha_q) $. If we identify $ f_q = f(\alpha_q) $ we can see that $ (q, \tau_q) $ and $ (\alpha_q, f_q) $ are related by a Legendre transformation
\begin{equation}\label{eq:legendre}
f_q = q \, \alpha_q - \tau_q \qquad \text{and} \qquad \alpha_q = \frac{\dif \tau_q}{\dif q} \, .
\end{equation}
It can be proven, e.g.\ in \cite{Janssen1994}, that $ \tau_q $ is a monotonically increasing function in $ q $, which implies that $ \alpha_q > 0, \forall q $.
\begin{figure}
	\centering
	\includegraphics[width=\textwidth]{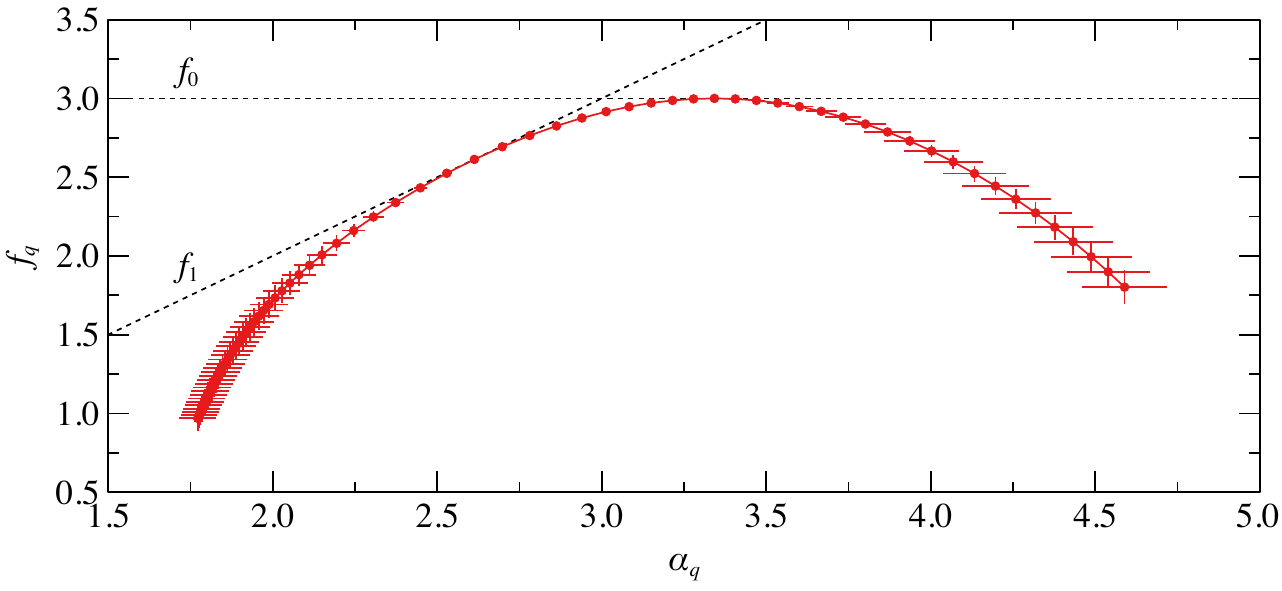}
	\caption{Multifractal spectrum $ f(\alpha) $ of the highest-occupied molecular orbital wave function of the \textsc{Onetep} prototype, computed for $ q $ from $ -2 $ to $ 5 $ in steps of $ 0.1 $ (increasing from right to left). Dashed lines indicate the functions $ f_0 \equiv D $ and $ f_1 (\alpha) = \alpha $. \label{fig:f_alpha_spectrum}}
\end{figure}
%
We can combine \emph{singularity strengths} $ \alpha_q $ and the \emph{singularity spectrum} $ f_q $ to obtain the \emph{multifractal spectrum} $  f(\alpha) $. This function is equivalent to the generalised dimensions $ D_q $ in characterising the multifractal, and in the case of a simple fractal analogously reduces to the point $ (D,D) $ in a $ (\alpha, f(\alpha)) $ plot.
%
As shown in the example of Fig.\ \ref{fig:f_alpha_spectrum}, $ f(\alpha) $ is a convex function reaching its maximum at $ \alpha_0 $ with a value $ f_0 = \tau_0 = D $. From \eqref{eq:legendre} we further notice that $ f_1 = \alpha_1 $, since $ \tau_1 = 0 $. The spectrum is therefore tangential to the functions $ f_0(\alpha) \equiv D $ and $ f_1(\alpha) = \alpha $.

\subsection{Symmetry of the multifractal spectrum}
Using the nonlinear $ \sigma $ model, Mirlin \emph{et al.} \cite{Mirlin2006a} have analytically proven that at criticality the multifractal exponents \eqref{eq:legendre} satisfy the exact symmetry relation
\begin{equation}\label{eq:mfe-symmetry}
\alpha_q + \alpha_{1-q} = 2 d \qquad f_{1-q} = f_q + d - \alpha_q \; .
\end{equation}
Assuming the universality of the critical properties at the Anderson transition, this result is expected to generally hold for the Wigner-Dyson symmetry classes \cite{Evers2008}. Indeed, Eq.\ \eqref{eq:mfe-symmetry} was confirmed numerically for different systems, including the 3D Anderson model \citep{Rodriguez2008,Vasquez2008a} and in experiments \citep{Faez2009}

\section{Multifractal analysis of the wave function}\label{sec:mfa-numerical}

We are mainly interested in the singularity strengths $ \alpha_q $, which, together with $ \tau_q $ and $ \Delta_q $ are called multifractal exponents (MFE). In this section we recast the exponents derived in Sec.\ \ref{sec:mfa} in a form that is more convenient for numerical calculations, mostly by reducing the loss of precision. We then extend our definitions to include a disorder ensemble average.

Before moving to the determination of these exponents, let us briefly discuss their expected values \cite{Vasquez2008a,Rodriguez2011}. In the case of very low disorder, the wave function intensities $|\psi(\mathbf{r})|^2$ will be nearly plane-wave like, i.e.\ extended throughout the whole of the volume $L^3$. Hence, as shown in section \ref{sec:mfa-measures}, $\alpha_q=D=d=3$ for all $q$. This implies that $\tau_q=d(q-1)$, $\Delta_q=0$ and $f_q=d$. Hence $f(\alpha)$ is seen to contract, and eventually converge, to one point $f(d)=d$.
On the other hand, for very large disorder, well into the insulating/localized regime, $|\psi(\mathbf{r})|^2$ can be seen as localized on a few (single) sites only. In this case $\alpha_q \rightarrow \infty$ for $q \leq 0$, and  $\alpha_q \rightarrow 0$ for $q > 0$. Similarly, $\tau_q \rightarrow -\infty$, $\Delta_q\rightarrow -\infty$ for $q \leq 0$, and  $\tau_q = 0$, $\Delta_q=d(1-q)$ for $q > 0$. The $f(\alpha)$ spectrum broadens and in the limit of strong localization, converges to the points $f(0)=0$ and $f(\infty)=d$.
At criticality, the behaviour is even richer \cite{Rodriguez2011}, with a non-parabolic $f(\alpha)$ \cite{Rodriguez2009} and $\alpha_0=4.048(4.045,4.050)$, $\alpha_1=1.958(1.953,1.953)$ the current best estimates for the non-interacting 3D Anderson model \cite{Rodriguez2011}. As a rough guide for the following sections, the $\alpha_q$ tend towards $d=3$ for extended states  as weak disorder, while $\alpha_0$ increases without bounds in the insulating regime.

\subsection{Numerical calculation}
Following \cite{Chhabra1989b}, it is convenient to define, from \eqref{eq:Rq} and \eqref{eq:legendre}, the auxiliary quantity
\begin{equation}\label{eq:Sq}
S_q(\lambda) = \frac{\dif R_q(\lambda)}{\dif q} = \sum_i \mu_i^q \log \mu_i \, .
\end{equation}
This ratio can be interpreted as an average with respect to the measure defined by $ \mu^q $. The latter is also called \emph{$ q $-microscope}, because it increases the large (small) box-probabilities for $ q > 0 $ ($ q < 0 $). A computationally-friendly formulation of the MFE reads
\begin{equation}\label{eq:mf-exp-num}
\tau_q = \lim_{\lambda \rightarrow 0} \frac{\log R_q (\lambda)}{\log \lambda} \, ,  \qquad
\Delta_q = \tau_q - q(d-1) \, ,  \qquad
\alpha_q = \lim_{\lambda \rightarrow 0} \frac{S_q(\lambda)}{R_q(\lambda) \log \lambda} \, .
\end{equation}
To comply with \eqref{eq:length-scales}, we choose $ \lambda \leq 1/2 $, namely we consider boxes of linear size up to $ l \leq L/2 $.
%
We coarse-grain the wave function using the partitioning scheme proposed by \cite{Thiem2013}. Here, the box size $ l $ can take any integer value (up to $ L/2 $), so that $ \lambda^{-1} = L/l $ can take non-integer values. This is achieved by first periodically replicating the original system, such that it can be exactly covered by an integer number of boxes, and then by averaging over the possible equivalent box origins. The increased number of available box sizes translates, in the linear fits, in reduced uncertainties in the estimated slopes.

\subsection{Ensemble averaging}
So far we have computed the multifractal properties of a single wave function. The multifractal analysis of the Anderson transition is usually performed by taking an average over the disorder realisations. The definitions of the MFE can be extended by defining the \emph{ensemble average} of the partition sum as $ \langle R_q(\lambda) \rangle \sim \lambda^{\tau_q^\text{ens}} $, such that
$
\tau_q^\text{ens} = \lim_{\lambda \rightarrow 0} {\log \langle R_q(\lambda) \rangle}/{\log \lambda} 
$. 
We then proceed to take the Legendre transform and define
\begin{equation}\label{eq:GMFEs-ensemble}
f_q^\text{ens} = q \alpha_q^\text{ens} - \tau_q^\text{ens} \qquad \text{and} \qquad
\alpha_q^\text{ens} = \frac{\dif \tau_q^\text{ens}}{\dif q} = \lim_{\lambda \rightarrow 0} \frac{\langle S_q(\lambda) \rangle}{\langle R_q(\lambda) \rangle \log \lambda} \, .
\end{equation}
Notice that, in the ensemble average of $ \alpha_q $, the $ q $-microscope $ \mu_i^q $ is normalised by $ \langle R_q \rangle $, namely the averaged partition sum of the wave function. If we normalised the $ \mu_i^q $ terms for every wave function we would obtain the \emph{typical} average $ \alpha_q^\text{typ} = \langle S_q/R_q \rangle/\log \lambda $ ($ \lambda \rightarrow 0 $). 
While in the ensemble average all wave functions, including rare events, are equally weighted, the typical average is dominated by the behaviour of ``typical'' wave functions.
The presence of rare events translates in the appearance of negative fractal dimensions (see Sec.\ \ref{ssec:neg-dimensions}), a feature of the $ f(\alpha) $ that is best captured by ensemble averaging \cite{Evers2008}.

\section{Results for eigenstates of the effective Hamiltonians}\label{sec:mfa-results}

For finite systems, the multifractal exponents are computed, as explained in Sec.\ \ref{sec:mfa-measures}, by estimating the slope of a $ \log \langle R_q (\lambda) \rangle $ vs.\ $ \log \lambda $ plot, in the case of $ \tau_q^\text{ens} $. Accordingly, the statistical uncertainties at fixed $ \lambda $ have to be multiplied by a factor $ \log \lambda $.%
\footnote{The standard deviation $ \sigma_{\alpha_q} $ associated to $ \alpha_q^\text{ens} $ (at a fixed $ \lambda $) is related to the standard deviations $ \sigma_{S_q} $ and $ \sigma_{R_q} $ and the covariance $ \cov (S_q,R_q) $ via propagation of the variance \citep{Rodriguez2011}
\begin{equation}
\sigma_{\alpha_q} = \alpha_q^\text{ens} \sqrt{\frac{\sigma^2_{S_q}}{\langle S_q \rangle^2} + \frac{\sigma^2_{R_q}}{\langle R_q \rangle^2} - 2 \frac{\cov (S_q,R_q)}{\langle S_q \rangle \langle R_q \rangle}} 
\quad \text{ and } \quad 
\label{eq:sigma-tauq}
\sigma_{\tau_q} = \frac{\sigma_{R_q}}{\langle R_q \rangle \log \lambda} \, ,
\end{equation}
so that $ \sigma_{f_q}^2 = \sigma_{\alpha_q}^2 + \sigma_{\tau_q}^2$. The standard error of the mean is obtained by dividing every standard deviation by $ \sqrt{\mathcal{N}} $, where $ \mathcal{N} $ is the number of available realisations.}
In Fig.\ \ref{fig:f_alpha_full} we show the average singularity spectrum for the ensemble of $ \num{10648} $ atoms with $ 140 $ impurities, a system that is critical at energy close to $ \SI{-0.249}{eV} $ \cite{Carnio2019}.
The increase in the ensemble size does not change the shape of the spectrum significantly, but has the effect of reducing the error bars on the data points.
In particular, it is the calculation of the average $ \langle R_q \rangle $ and $ \langle S_q \rangle $ in Eq.\ \eqref{eq:GMFEs-ensemble} that benefits from larger ensembles, since smaller error bars in the data used for the linear fits results in smaller uncertainties on the fit parameters, see Fig.\ \ref{fig:MFA-linear-fits}(b).
We quantitatively report the quality of said fits in the lower panel of Fig.\ \ref{fig:f_alpha_full}(a), where we show the linear correlation coefficient $ r^2 $ and the $ p $ value. As noted in \cite{Rodriguez2008}, while $ r^2 \approx 1 $ indicates a good linear behaviour, small $ p $ values suggest that the uncertainties on the data point are too small to support the deviation from the linear behaviour we are fitting. This is likely due to the limited number of realisations available for the ensemble averaging.
%
For comparison, at the end of the two branches, i.e.\ for large $ |q| $ values, error bars are larger and hence the quality-of-fit increases again.

\subsection{Negative fractal dimensions}\label{ssec:neg-dimensions}
Error bars increase on the two ends of the spectrum, for negative $ q $ (right) and positive (left). For $ q < 0 $, the $ q $-microscope increases the weight of small values of the wave function, which are more sensitive to numerical fluctuations from the diagonalization. The other end of the spectrum ($ q > 0 $, left) describes instead the presence of \emph{rare} critical wave functions with small values of $ \alpha $ and hence large $ |\psi|^2 \sim L^{-\alpha}$ \cite{Rodriguez2009}. The set of these values scales with a negative fractal dimension $ f(\alpha) $, which means that their occurrence frequency vanishes in the $ L \rightarrow \infty $ limit. This is a known effect arising from ensemble averaging \citep{Chhabra1991} and known since the pioneering work of Mandelbrot \citep{Mandelbrot1984}.
This finite-size effect is further observed and commented in \cite{Rodriguez2008}, where larger systems are accessible and studied. In comparison, the single state used to produce Fig.\ \ref{fig:f_alpha_spectrum} does not show said rare boxes with large probability amplitudes, as indicated by $ f_q > 0 $.

\subsection{Width of the multifractal spectrum}
Let us comment on the width of the distribution in Fig.\ \ref{fig:f_alpha_full}, as compared to the Anderson model studied in \cite{Rodriguez2008}.
A narrow $ f(\alpha) $ spectrum implies that extreme values (either large or small) occur less frequently. This means that, in our case, the average state near criticality in our model looks more homogeneous or extended than in the Anderson model.
%
In Fig.\ \ref{fig:fa-energies}, instead, we show the singularity spectrum for a system of $ \num{10648} $ atoms, which, for $ 230 $ impurities, is close to criticality at the Fermi energy $ \varepsilon_\text{F} = 0 $ and deeper in the impurity band at $ \SI{-0.320}{eV} $ (estimated from \cite{Carnio2019}). While both spectra are narrower than the Anderson model, the critical wave function at $ \varepsilon_\text{F} $ appears on average more extended than deeper in the impurity band. 

%
Our results are reminiscent of those found by \cite{Mirlin2000b} for the power-law random banded matrix (PRBM) model, which describes a 1D chain with random long-range hopping decaying as $ r^{-\alpha} $ over distances larger than a band width $ b $.
For the critical value $ \alpha = 1 $, the model undergoes an Anderson transition for any value of $ b $, which parametrises a family of critical models
%
that can be studied from the weak- ($ b \gg 1 $) to the strong-coupling ($ b \ll 1 $) regime.
For $ b \gg 1 $ the model shows a ``quasi-metallic'' behaviour, where the critical wave functions has statistical properties similar to the delocalised phase. The singularity spectrum becomes correspondingly narrower with a parabolic shape, a regime called \emph{weak multifractality}. In this case the multifractal spectrum follows the parabolic \emph{approximation} \citep{Janssen1994}:
\begin{equation}\label{eq:weak-mf}
f(\alpha) \simeq d - \frac{(\alpha - \alpha_0)^2}{4(\alpha_0-d)} \qquad \text{and} \qquad \alpha_0 = d + \gamma \, .
\end{equation}
In Fig.\ \ref{fig:f_alpha_full} we fit our full-ensemble data to \eqref{eq:weak-mf} to find the estimate value 
$ \alpha_0 \approx 3.55 $. We only report an approximate value without uncertainty because the parabolic behaviour is an approximation and does not necessarily hold for the whole spectrum, since $ f(\alpha) $ is defined only for positive $ \alpha $. In fact, for $ q_c = (d+\gamma)/2\gamma $ \citep{Evers2008} we have $ \alpha_{q_c} = 0 $ and $ f(0) $ is finite (a \emph{termination point}), whereas our results in Fig.\ \ref{fig:f_alpha_full} and those in \cite{Rodriguez2008} seems to suggest that $ f(\alpha) \rightarrow - \infty $ in the limit $ \alpha \rightarrow 0 $ (hence no termination point).

\subsection{Symmetry of the multifractal spectrum}
\begin{figure}
	\centering
	\includegraphics[width=\textwidth]{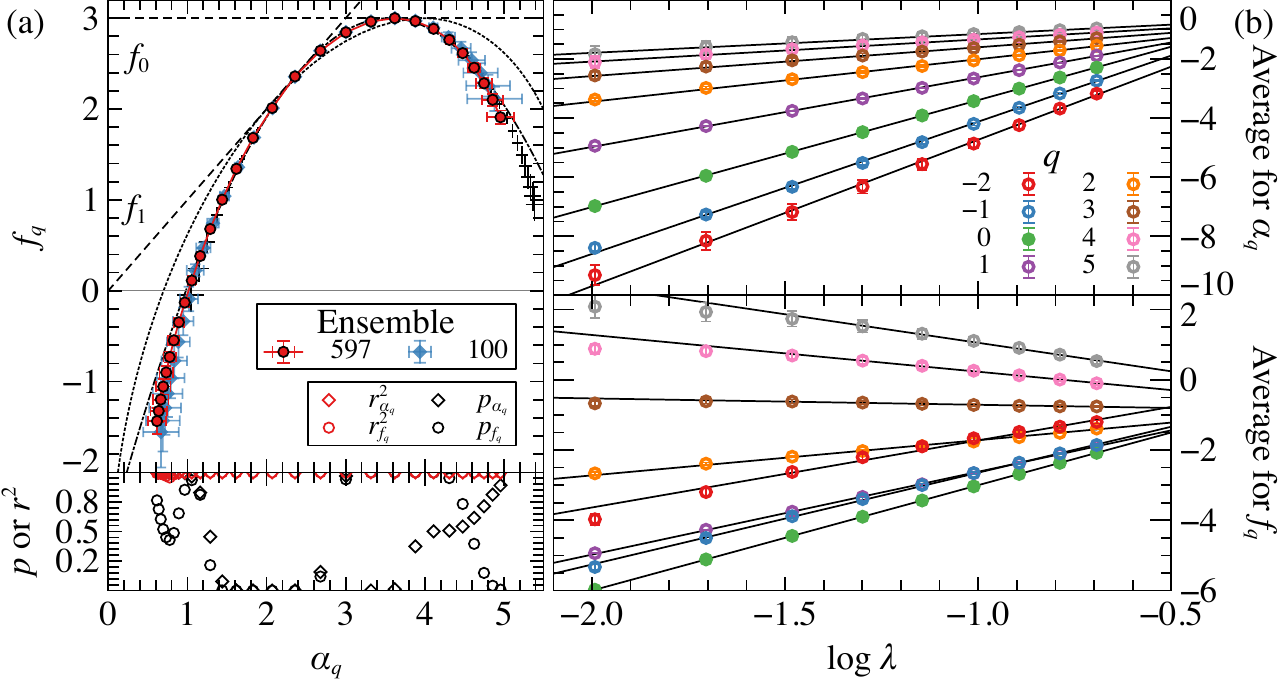} \hfill
	\caption{
	(a) top panel: singularity spectrum for $ L = 22 $ and $ N_\text{S} = 140 $, sampled for values of $ q $ from $ -2 $ to $ 5 $ in steps of $ 1/4 $ (increasing from right to left) at energy $ \SI{-0.249}{eV} $. Blue diamonds show the results for the ensemble of the first 100 disorder realisations, while red circles indicate the results from all available realisations (597). Simple error bars, without data point, indicate the symmetrised spectrum to the full ensemble. Dashed lines indicate the functions $ f_0 \equiv D $ and $ f_1 (\alpha) = \alpha $. The dotted line indicates the spectrum for the Anderson model at criticality, reproduced from \cite{Rodriguez2008}, while the dot-dashed line indicates the fit to the parabolic approximation \eqref{eq:weak-mf}. \label{fig:f_alpha_full}
	(a) bottom panel: linear correlation coefficient $ r^2 $ (red) and quality of fit $ p $ (black) for the linear fits used to extrapolate the thermodynamic limit of $ \alpha_q $ (diamonds) and $ f_q $ (circles), shown in Fig.\ \ref{fig:MFA-linear-fits}.
(b)
		Linear fits used to produce the data plotted in Fig.\ \ref{fig:f_alpha_full}. The slopes of the lines yield $ \alpha_q $ (panel above) and $ f_q $ (below). For clarity we only show data for integer values of $ q $ from $ -2 $ (red) to $ 5 $ (grey), with data for $ q = 0 $ highlighted with a full symbol. \label{fig:MFA-linear-fits}}
\end{figure}

\begin{figure}
	\centering
	\includegraphics[width=\textwidth]{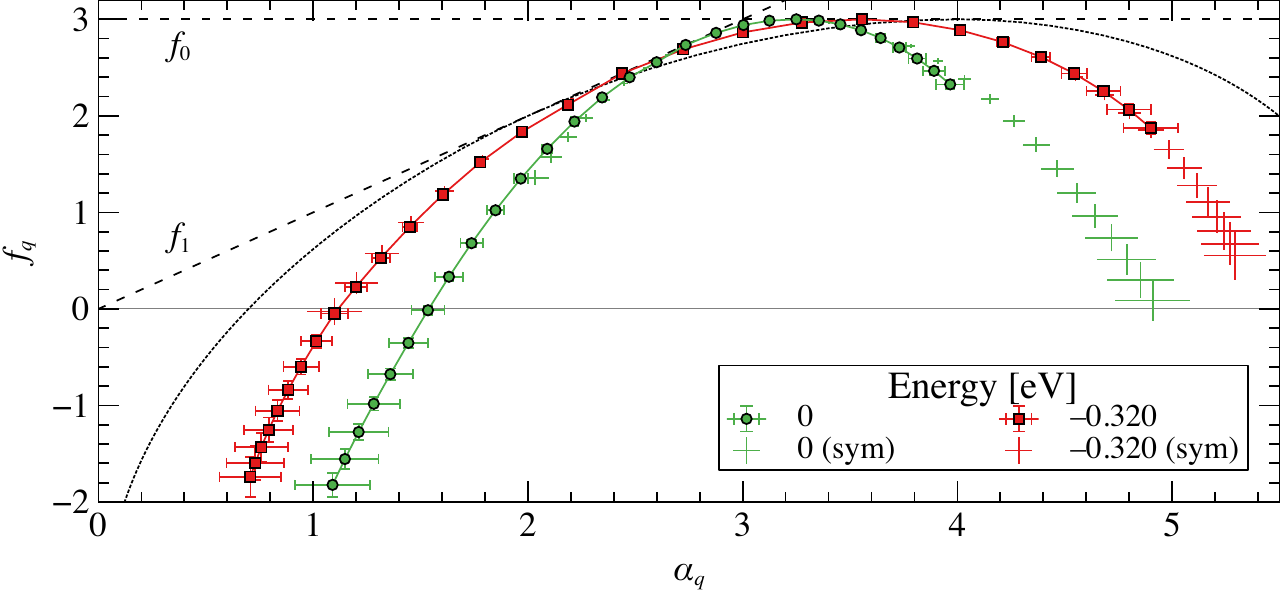}
	\caption{Singularity spectrum for $ L = 22 $ and $ N_\text{S} = 230 $, sampled for values of $ q $ from $ -2 $ to $ 5 $ in steps of $ 1/4 $ (increasing from right to left). The ensemble contains $ 500 $ realisations. Green circles indicate the average over the ensemble of states near the Fermi energy $ \varepsilon_\text{F} = 0 $, while red squares indicate the ensemble over states closest to $ \SI{-0.320}{eV} $. The corresponding symmetrised spectra are indicated with the same colours by the error bars only. \label{fig:fa-energies}}
\end{figure}

In Fig.\ \ref{fig:f_alpha_full} we also show the symmetrised spectrum obtained by computing and plotting $ \alpha_{1-q} $ and $f_{1-q}$ from Eq.\ \eqref{eq:mfe-symmetry}. As expected from the previous paragraphs, the uncertainty on the data points increase at the extremities. Within these error bars, the spectra are in good agreement with each other.
%
%
We verify the same symmetry relation also for the spectra in Fig.\ \ref{fig:fa-energies}.
While at $ \SI{-0.320}{eV} $ there is excellent agreement between the spectra, at the Fermi energy there is a slightly higher discrepancy, especially at extreme values of $ q $. Since this discrepancy would be resolved by taking two standard deviations as confidence intervals, instead of one, we cannot attribute this discrepancy to any specific underlying physical factor or systematic error.

\section{Validity of the scaling assumption}
\label{sec-scaling}

The question that arises when dealing with finite systems is whether the wave function is still a multifractal. Formally speaking, the wave function is a true multifractal only at the critical point. For a finite $ L $, however, the effective critical point shifts away as $ L^{-1/\nu}$ from its thermodynamic limit \citep{Cardy1996}.
%
Luckily this is not a problem, since, as shown by \cite{Cuevas2007}, states on the two sides of the transition still show multifractal features characteristic of a critical wave function.

\begin{figure}[ptb]
	\centering
	\includegraphics[width=\textwidth]{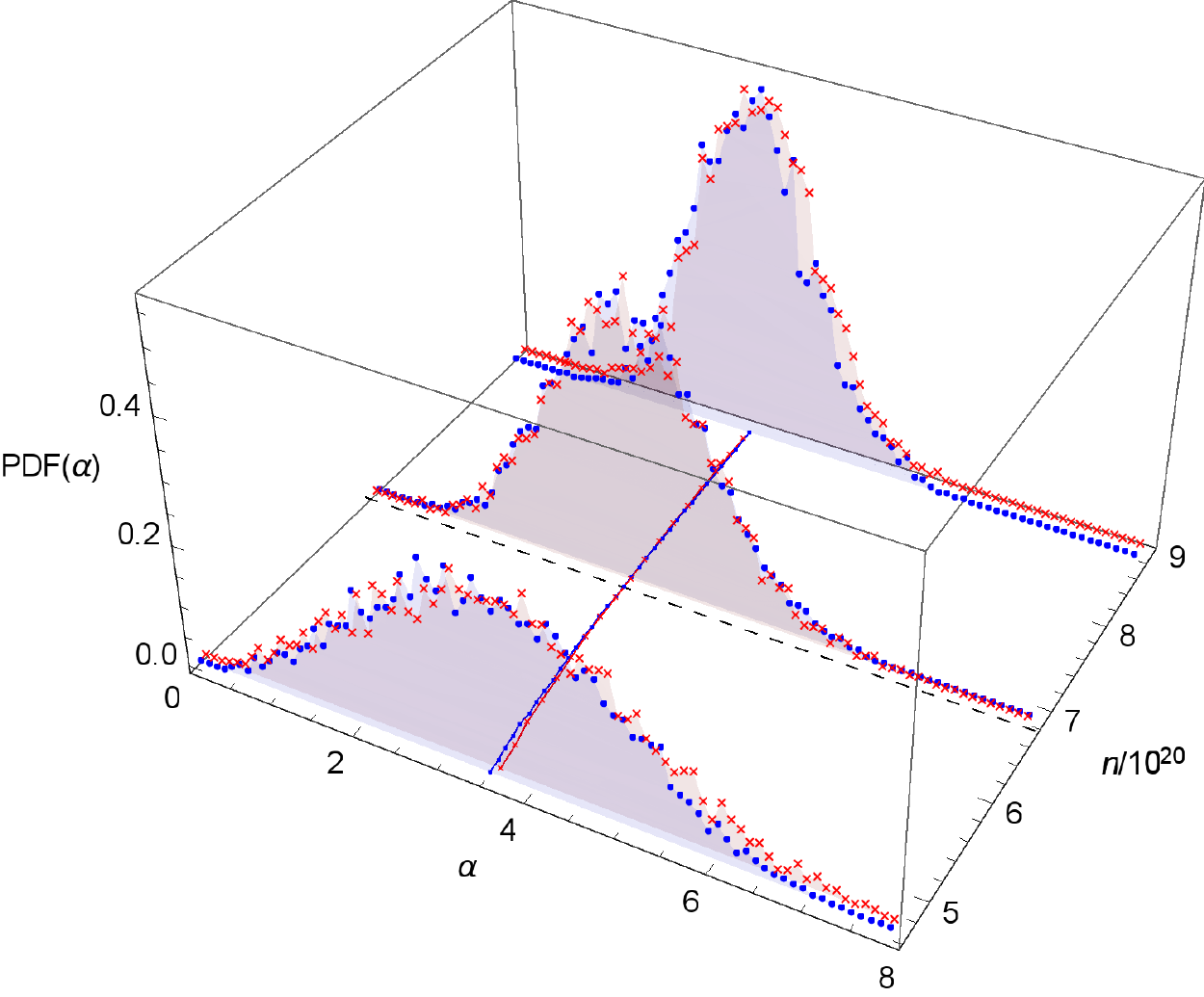}
	\caption[System-size invariance of $\PDF(\alpha)$ at the Anderson transition.]{Ensemble PDF of the multifractal measure $ \alpha $ at coarse-graining $ \lambda = 1/2 $ and energy $ \varepsilon - \varepsilon_\text{F} = \SI{-0.249}{eV} $, as a function of the concentration $ n $ (in units of $ \SI{E20}{cm^{-3}} $), for two system sizes $ L^3=4096 $ (blue dots) and $ 10648 $ (red crosses). For clarity we show the histogram for three concentrations: before the transition ($ n = \SI{4.6E20}{cm^{-3}} $), near the critical point ($ \SI{6.8E20}{cm^{-3}} $), and after ($ \SI{8.8E20}{cm^{-3}} $). The critical point ($ n_\text{c} = \SI{6.7E20}{cm^{-3}} $) is indicated by a black dashed line \cite{Carnio2019}. On the bottom plane we show the position of the average $ \alpha_0 $ also for the intermediate concentrations, connected by lines to guide the eye. We use again blue dots with a solid line for $ L^3=\num{4096} $ and red crosses with a dashed line for $ 10648 $. Reproduced from the Supplemental Materials to \cite{Carnio2019}.}
	\label{fig:pdf-invariance}
\end{figure}

\begin{figure}[tb]
	\centering
	\includegraphics[width=\textwidth]{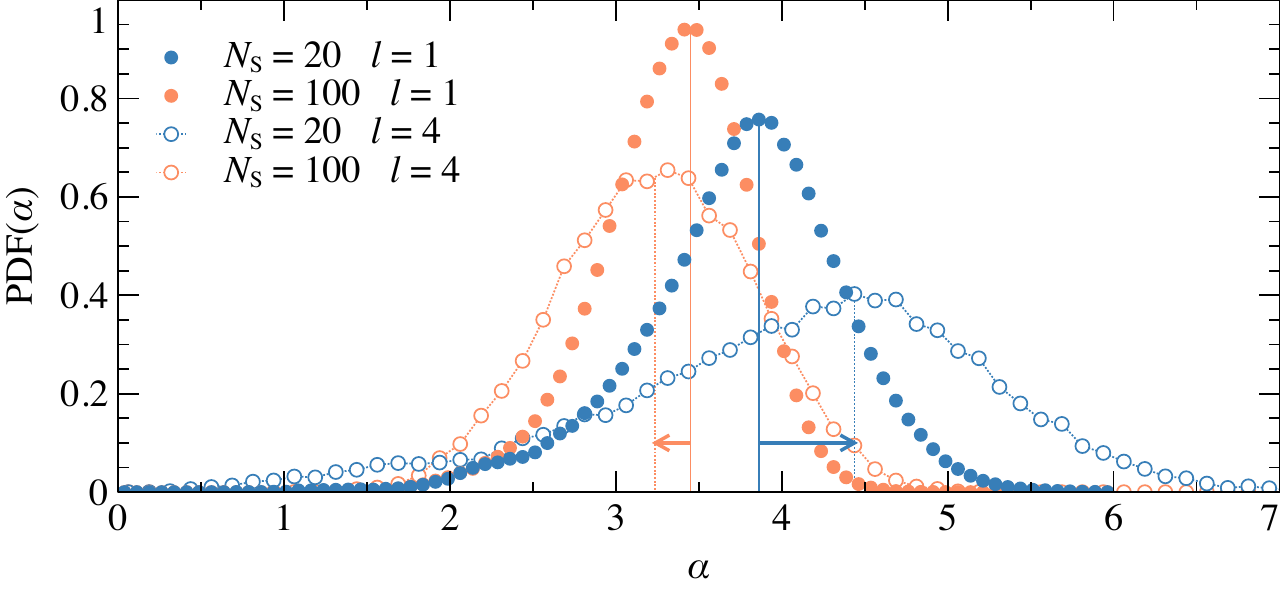}
	\caption{Ensemble PDF of the multifractal measure $ \alpha $ at energy $ \varepsilon - \varepsilon_\text{F} = \SI{-0.249}{eV} $ for $ L^3=4096 $ atoms. In blue we show the PDF for $ N_\text{S} = 20 $, in orange for $ N_\text{S} = 100 $. Filled symbols indicate a coarse-graining with box size $ l = 1 $, while empty symbols connected by a dotted line indicate $ l = 4 $. The thin vertical lines indicate the average values of $\alpha_0$ for each case. 
	}
	\label{fig:pdf-alphas}
\end{figure}

Now that we can construct a multifractal measure from the wave functions away from the critical point, we can actually check the most important assumption we have taken so far, namely that a localisation-delocalisation transition occurs in our model.
The histogram distribution $ N_\lambda(\alpha) $ of the measure $ \alpha $ depends, at the critical point, only on the coarse-graining $ \lambda = l/L$, rather than separately on the system size $ L $ and the box size $ l $. 
At the critical point, then, $ N_\lambda(\alpha) $ has the same shape for any $ L $, provided that the wave functions are coarse-grained with the matching $ l $ box size. 
The dependence of $ N_\lambda(\alpha) $ on $ L $ gradually reappears away from the critical point, where in the strong (weak) disorder regime, larger systems become more localized (delocalized). 
This is shown in Fig.\ \ref{fig:pdf-invariance}, where we plot the ensemble histogram $ \PDF(\alpha) = N_\lambda(\alpha) \lambda^d / \mathcal{N} $ at $ \lambda=1/2 $ for three values of $ n $: the lowest in the localized regime, the intermediate close to the critical point and the highest in the delocalised regime. 
The ensemble $ \PDF(\alpha) $ is built by filling one single histogram with the $ \mathcal{N} $ available wave functions.
%
Because of the limited spread in system sizes and the large common coarse-graining, the difference in the PDFs outside the critical point is not very well pronounced in Fig.\ \ref{fig:pdf-invariance}. An alternative check consists in fixing the system size and study how the PDF renormalises with the coarse graining \citep{Lindinger2017}.
The ``scaling'' variable $ \xi / L $ \citep{MacKinnon1981} scales like $ \lambda^{-1} $, which implies that, with increasing $ \lambda $, $ \xi/L $ becomes smaller. Physically this means that, upon coarse-graining, localised (delocalised) states become more localised (delocalised), or, equivalently, that the renormalisation flow rescales the disorder away from its critical value,
if a phase transition, and hence a critical point, exists \cite{Carnio2018}. We verify this in Fig.\ \ref{fig:pdf-alphas} where the PDF's move in opposite directions upon increasing the box size $ l $ in a system of $ L^3 = \num{4096} $ atoms.

\section{Conclusions}
\label{sec:conclusions}

After reviewing the foundations of multifractal analysis in the study of a disordered system at criticality, we have shown the multifractal nature, captured by the $f(\alpha)$ spectrum, of the Kohn-Sham wave function. This result is in line with previous experimental \citep{Richardella2010}, theoretical \citep{Burmistrov2013}, and numerical \citep{Amini2014,Harashima2014,Lee2018} studies, where the critical fluctuations of the wave function at criticality are expected to survive in the presence of the Coulomb interaction.

We have shown that multifractality persists also in the wave functions near the critical concentrations computed in the effective tight-binding model presented in Refs. \cite{Carnio2019,Carnio2018}.  The multifractal spectrum follows the symmetry relations derived from field-theoretical models \citep{Mirlin2006a} and the ensemble average over hundreds of realizations results in negative fractal dimensions ascribed to rare events. These are known features of (finite) multifractal critical states that have been studied also in the non-interacting 3D Anderson model  \cite{Rodriguez2010}. A difference with this case, however, lies in the nearly-parabolic shape of the $f(\alpha)$ spectrum observed here. This behaviour, referred to as weak multifractality, is a common trait with the Anderson transition in $ 2+\epsilon$ deminsions, with $ \epsilon \ll 1 $, and with the power-law random banded matrix model with $ b \gg 1 $ \citep{Evers2008}.

Finally, we compute the $ f(\alpha) $ spectrum near the two mobility edges, respectively, far from and at the Fermi energy. In the latter case we notice that the spectrum is narrower, indicating that the critical wave function at this energy is, on average, more delocalized than deeper in the band. In  Refs.\ \cite{Carnio2019,Carnio2018} we propose the origin of this quasi-metallic behaviour to be the hybridization of the impurity states near the Fermi energy with states from the conduction band. The presence of this second extended band might alter the physics of the metal-insulator transition close to the Fermi energy, leading to a varying critical exponent across the impurity band.

A similar phenomenon has very recently been reported in a Hartree-Fock study of the Anderson transition in the presence of tunable Coulomb interaction \cite{Lee2018}. There the authors show that a second mobility edge appears near the Fermi energy, where the Coulomb gap forms. At this energy, the critical state shows a narrower multifractal spectrum compared to the mobility edge at higher energy. While in this model quasi-metallic behaviour appears where the Coulomb gap forms in the centre of the band of the Anderson model, it is present in our model when the impurity band forms and then merges with the conduction band.

To conclude, the numerical analysis of the Anderson transition in the presence of interactions, and, in particular, in real materials, is still very challenging. Progress relies on a large amount of resources needed for the simulations, with recent first-principles or self-consistent calculations reaching system sizes of the order of $10^2$--$10^4$ sites, and of hundreds of realisations \cite{Carnio2019,Lee2018,Zhang2018a} --- while studies of the non-interacting Anderson model can nowadays easily achieve more then $10^6$ sites and $10^4$ samples \cite{Rodriguez2011,Lindinger2017}. Still, \textit{ab initio} studies of real materials have observed new phenomena and traced these back to the electronic interaction. Future investigations with increasing system and sample sizes will undoubtedly clarify the universal properties of the transitions and critical points.




\section*{Acknowledgements}

RAR thanks Presidency University, Kolkata, where part of this work was written, for their warm hospitality. We thank EPSRC for support via the ARCHER RAP project e420 and the MidPlus Regional HPC Centre (EP/K000128/1). UK research data statement: data is available at 
{http://wrap.warwick.ac.uk/id/eprint/92910}
.





\bibliographystyle{prsty}
\bibliography{Anderson+DFT.bib}







\end{document}